# Highly-efficient spintronic terahertz emitter enabled by metal-dielectric photonic crystal

Zheng Feng[1], Rui Yu[2], Yu Zhou[3], Hai Lu[4], Wei Tan[1,*], Hu Deng[5], Quancheng Liu[5], Zhaohui Zhai[6], Liguo Zhu[6], Jianwang Cai[3,*], Bingfeng Miao[2], and Haifeng Ding[2,*]

[1] *Microsystem & Terahertz Research Center, CAEP, Chengdu 610200, China*

[2] *National Laboratory of Solid State Microstructures and Department of Physics, Nanjing University, Nanjing 210093, China*

[3] *Beijing National Laboratory for Condensed Matter Physics, Institute of Physics, Chinese Academy of Sciences, Beijing 100190, China*

[4] *Engineering Laboratory for Optoelectronic Technology and Advanced Manufacturing, Henan Normal University, Xinxiang 453007, China*

[5] *School of Information Engineering, Southwest University of Science and Technology, Mianyang 621010, China*

[6] *Institute of Fluid Physics, China Academy of Engineering Physics, Mianyang 621900, China*

*Corresponding author: *tanwei@mtrc.ac.cn, jwcai@iphy.ac.cn, hfding@nju.edu.cn*

## Abstract

Spintronic terahertz (THz) emitter provides the advantages such as apparently broader spectrum, significantly lower cost, and more flexibility in compared with the commercial THz emitters, and thus attracts great interests recently. In past few years, efforts have been made in optimizing the material composition and structure geometry, and the conversion efficiency has been improved close to that of ZnTe crystal. One of the drawbacks of the current designs is the rather limited laser absorption - more than 50% energy is wasted and the conversion efficiency is thus limited. Here, we theoretically propose and experimentally demonstrate a novel device that fully utilizes the laser intensity and significantly improves the conversion efficiency. The device, which consists of a metal-dielectric photonic crystal structure, utilizes the interference between the multiple scattering waves to simultaneously suppress the reflection and transmission of the laser, and to reshape the laser field distributions. The experimentally detected laser absorption and THz generations show one-to-one correspondence with the theoretical calculations. We achieve the strongest THz pulse emission that presents a 1.7 times improvement compared to the currently designed spintronic emitter. This work opens a new pathway to improve the performance of spintronic THz emitter from the perspective of optics.

Terahertz (THz) radiation plays increasingly important roles in both scientific research and practical applications, such as material science [1–3], biomedicine [4–6], wireless communication [7], and security imaging [6,8,9] etc. In all aspects, it is critical to have a wide band, fully controllable THz source with high power. Despite its high importance, the progresses of high-performance THz devices, especially the emitters [10,11], are still lacking behind. This is in sharp contrast to the well-developed technologies in its neighboring infrared and microwave bands, leaving a gap so called 'terahertz gap'. To date, the development of the THz emitters with high power and high efficiency is still one of the foci in the field of THz research [12,13].

 The femtosecond laser driven THz emitter is an important type of THz emitters that is commonly used [2,6,11]. Previously, its generation was mainly based on the non-spin mechanisms, such as the transient electrical currents in photoconductive antennas [14–18], the optical rectification from electro-optical crystals [19–24], and the plasma oscillations [25,26] etc. In 2013, T. Kampfrath *et al*. demonstrated a new type THz emitter, called spintronic THz emitter, which is based on the spin related effects in ferromagnetic/non-magnetic (FM/NM) heterostructures [27]. Compared to the conventional non-spin based THz emitters, the spintronic THz emitter possesses the additional advantage of the spin freedom of the electron besides that of the charge freedom, opening a new pathway for broadband (up to 30 THz) and controllable THz wave generation [27]. The reported conversion efficiency of the spintronic THz emitter, however, was about two orders of magnitude lower than that of commercial ZnTe crystal. Since then, efforts have been made to improve the performance [28–30]. In 2016, T. Seifert *et al*. significantly enhanced the conversion efficiency by choosing NM layer with large spin Hall angle (such as Pt, W), optimizing the layers' thickness as well as introducing $NM_1$/FM/$NM_2$ trilayer to fully utilize both the backward- and forward-flowing spin current. They found that a 5.8-nm-thick W/CoFeB/Pt trilayer achieved the conversion efficiency close to that of the commercial 1-mm-thick ZnTe crystal and outperform it in terms of the bandwidth, flexibility, scalability and cost [28]. D. Yang *et al*. proposed the cascaded multilayer which can generate transient current in each Pt layer, leading to considerable power increase [29]. Y. Wu *et al*. made a comprehensive study on FM/NM bilayer structures, and showed the capability of fabrication on flexible substrates [30]. These optimizations cover from the perspective of both spintronics and THz pulse generation [28–30], and furthermore, THz beam focusing has been utilized to obtain higher intensity [31,32], yet the conversion efficiency remains unchanged. On the other hand, the spintronic emitter also involves the laser pumping and absorption processes, which has not been explored yet. In particular, in previous designs [28–32], the ultra-thin metal films only absorb rather limited laser intensity and more than half of the intensity is wasted. To reach higher conversion efficiency, this drawback must be overcome.

Here, we demonstrate a novel scheme to improve the performance of the spintronic THz emitter in turns of power intensity. It utilizes the metal ($NM_1$/FM/$NM_2$)-dielectric photonic crystal (PhC) structure, where the multiple scatterings suppress the reflection and transmission of the laser light simultaneously, thus maximizing the laser field strength in the metal layers. Since the dielectric interlayer is almost non-dissipative, most of the laser energy absorption occurs in the $NM_1$/FM/$NM_2$ heterostructure, which improves the conversion efficiency of the spintronic emitter. The idea is first presented theoretically with the transfer matrix method based model. We further experimentally fabricate a series of PhC samples with different periods and repeats. The measured laser absorbance and the THz amplitude show one to one correspondences with the theoretical calculations. At the optimal conditions, the experimentally obtained conversion efficiency of the photonic crystal structures is about 1.7 times as high as that of the single-repeat spintronic THz emitter, demonstrating the validity of the proposed method. Our work opens a new pathway to improve the performance of the optically pumped spintronic emitter from the perspective of optics.

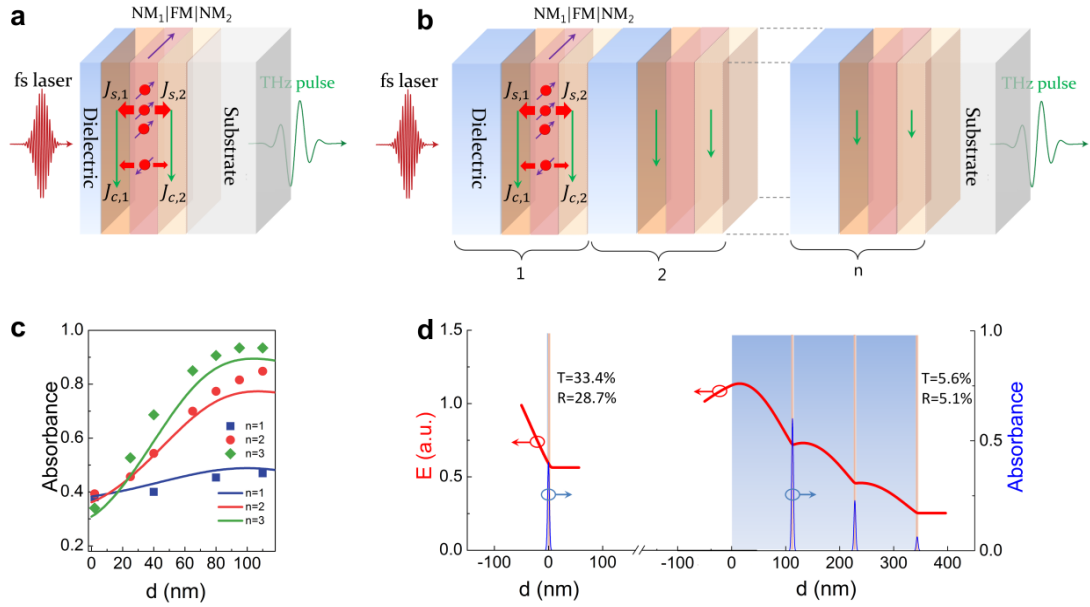

**Figure 1.** a) Schematic of single-repeat spintronic THz emitter. b) Schematic of the metal-dielectric photonic crystal type spintronic THz emitter. c) The $SiO_2$ thickness dependent femtosecond laser absorbance for n=1,2,3. Symbols are experimental results and solid lines are theoretical calculations. d) The calculated laser field distribution (red) and absorbance in metal layers (blue) of the single-repeat and metal-dielectric photonic crystal type emitter.

Spintronic THz emitter relies on two fundamental spintronic effects: the ultrafast laser pulse induced spin polarized current generation and the conversion of the spin current to charge current, namely the inverse spin Hall effect [33,34]. **Figure 1**a shows the schematic processes of a typical spintronic THz emitter with a trilayer heterostructure consisting of $NM_1$/FM/$NM_2$ thin films：(i) a femtosecond laser pulse impinges on the heterostructure and pumps ultrafast spin currents ($J_{s,1}$ and $J_{s,2}$) from the FM layer into the adjacent $NM_1$ and $NM_2$ layers; (ii) due to the inverse spin Hall effect, the ultrafast spin currents are converted into transient charge currents ($J_{c,1}$ and

$J_{c,2}$) along the *y* direction, leading to the THz emission out of the structure. In order to maximize the THz emission, $J_{c,1}$ and $J_{c,2}$ should be parallel to form unidirectional charge current, which requires the two NM layers to have spin Hall angles with opposite sign. Pt and W have been proved to be a good choice [28].

In previous works [28–30], great efforts have been made to maximize the emission efficiency via optimizing the thickness of FM and NM layers. The total thickness of either FM/NM bilayer structure or $NM_1$/FM/$NM_2$ trilayer structure is suggested to be less than 6 nm [28–30]. This thickness is smaller than the skin depth of the metallic heterostructure and significantly smaller than the laser wavelength (800 nm). As a result, more than half of the incident laser energy is either reflected by or transmitted through the sample, which strongly limits the conversion efficiency. We present a toy model here to offer the evidence. For simplicity, but without losing the generality, we assume that (i) the thickness is negligible compared to the wavelength, and (ii) the dielectric constant of the environment is uniform. According the boundary conditions and the Maxwell's equations, the reflection and transmission coefficients should satisfy the relation of $1 + r = t$. Consequently, one can deduce that the absorbance, $A = 1 - |r|^2 - |t|^2$, to be ≤50% (see details in Supporting Information Note 1). Note that this is an intrinsic limitation to the absorbance of ultra-thin metal films. To overcome this limit, we propose to use the metal-dielectric PhC structure, where the multiple scatterings and interferences could efficiently suppress the reflection and transmission simultaneously, thus the laser absorption in the metal films is significantly increased.

The schematic of the proposed structure is shown in Figure 1b. It is composed of periodic metal-dielectric films, [dielectric interlayer/NM$_1$/FM/NM$_2$]$_n$, on MgO substrate, where n denotes the number of repeats. In this work, we choose Pt(1.8 nm)/Fe(1.8 nm)/W(1.8 nm) for the THz emitter, which exhibits the largest THz emission efficiency in our experiments, and SiO$_2$ for the dielectric interlayer. Multiple scatterings and interferences in such structure can be tailored by adjusting the thickness of the period (*d*) and the number of repeats (n). Transfer-matrix method is employed for the theoretical calculations and the structure design, which has been proved to be an efficient tool [28,35].

Since the thickness of each metal film is smaller than its skin depth, the permittivity is most likely different with their bulk values. Thus, we treat the Pt(1.8 nm)/Fe(1.8 nm)/W(1.8 nm) structure as a "single layer" and retrieve its effective permittivity from the measured reflection and transmission coefficients. The model fitting showed that the effective permittivity to be $\varepsilon_m$=−29.36+24.01$i$ (see Supporting Information Note 3). In the calculation, we also used the bulk value of the permittivity for SiO$_2$ interlayer and MgO substrate, namely, 2.11 for SiO$_2$ [36] and 2.98 for MgO [37].

The calculated absorbance, defined as $A = 1 - |r|^2 - |t|^2$, as a function of the thickness of SiO$_2$ interlayer for a series of samples, n=1,2,3, is plotted as solid lines in Figure 1c. One can see that for n=1, the absorbance slightly increases with increasing thickness and reaches a maximum at about *d*=100 nm, yet the maximum absorbance is below 50%. For n=2 and 3, the absorbance increases sharply with increasing thickness of SiO$_2$ interlayer, and the maximum value increases with the increase in n. It is worth noting that the maximum absorbance for n=3 is almost 90%, about twice of the maximum value for n=1. In contrast, when *d* is smaller than 25 nm, the absorbance shows inverse relation with n, which decreases with increasing n. This can be intuitively understood that the dielectric interlayers are too thin to modify the interference of scattering waves and the increase of n is approximately equivalent to the increase of the thickness of metal films. Hence, the reflection is considerably enhanced, leading to a reduced absorption.

To gain further insight into the interference in periodic structures, we depict in Figure 1d the laser field distribution in two samples: one with n=1 and *d*=2 nm, which was firstly realized in Ref. [28], and the other one with n=3 and *d*=110 nm, according to our theoretical design. For the n=1 sample, the field decays exponentially in the metal films, and the reflectance ($R = |r|^2$) and transmittance ($T = |t|^2$) are calculated to be 28.7% and 33.4%, respectively. For the n=3 sample, the laser field distribution is reshaped to suppress the reflection and transmission simultaneously, where $R = 5.1\%$ and $T = 5.6\%$. The laser field in the dielectric interlayers exhibits a waveform-like profile, resulting from the interference of scattering waves. In comparison with the n=1 sample, the laser field in the first stack of metal films is "lifted", and extended into the second and third stacks, indicating that more incident energies are trapped and absorbed. Therefore, each period can act as an emitter. Since the total thickness of the n=3 sample is less than 350 nm, far smaller than the quarter wavelength of the emitted THz radiation, the induced transient charge currents in each period are in phase, leading to the constructive interference of all emitters. It is expected that the greatly enhanced laser absorption and the superposition of THz emissions would significantly improve the conversion efficiency of the spintronic THz emitter.

A series of samples with n=1,2,3 and $d$ varying from 2 nm to 110 nm were fabricated to verify our theoretical proposal. We firstly measured the reflectance and transmittance of the samples under fs laser illumination to obtain their absorbance. The blue squares, red circles, and green diamonds in Figure 1c depicts the fs laser absorbance as a function of $SiO_2$ thickness $d$ with different repeating periods, n=1,2,3, respectively. The experimental results exhibit excellent agreement with the theoretical calculations (solid lines). The highest absorbance of the samples with n=2 and n=3 reaches 82% and 93%, respectively. Compared to the single-repeat spintronic emitter, for example, the one with n=1 and $d$=2 nm, the absorbance in the photonic crystal structures is enhanced by a factor of more than 2.

We then continue with the investigation of the enhancement of THz conversion efficiency. **Figure 2**a–c show the measured THz pulses generated by the spintronic emitters as a function of $SiO_2$ thickness $d$ for different repeats, n=1,2,3. To show the effect quantitatively, we define the THz pulse amplitude as the peak to peak intensity and plot its $d$-dependence for different n in Figure 2d. It is readily to find that the THz pulse amplitude increases with $d$ for each period, in accordance with the trend of fs laser absorbance. Nevertheless, the situation becomes quite different if we compare the samples with the same $d$ but different n. For example, when $d$=2 nm, the THz pulse amplitude of the emitter with n=2 is obviously smaller than that of n=1, whereas they have comparable laser absorbance. When $d$>40 nm, although the laser absorbance of n=2 is smaller than that of n=3, the emitted THz pulse amplitude is considerably larger. The sample with n=2 and $d$=110 nm provides the strongest THz pulse emission, which present a 1.7 times improvement compared to the single-repeat spintronic THz emitter with $d$=2 nm.

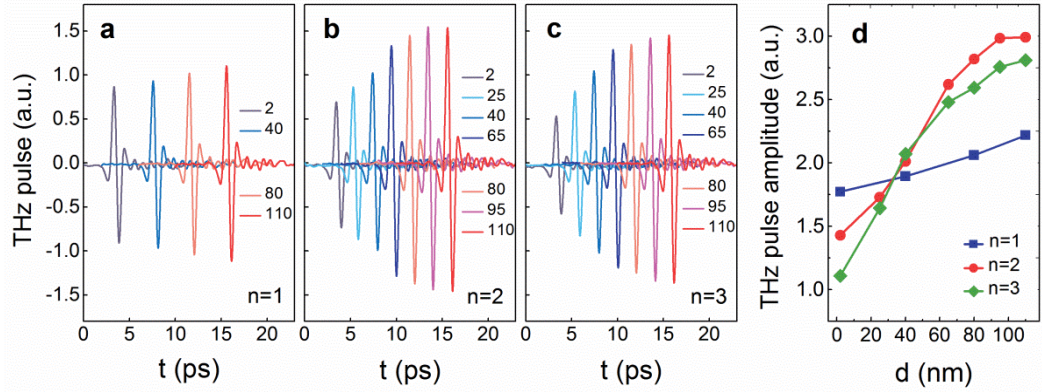

**Figure 2.** a–c) Experimentally measured $d$-dependent THz pulse generation from samples $[SiO_2(d)/Pt(1.8\ nm)/Fe(1.8\ nm)/W(1.8\ nm)]_n$ with different repeats, namely, n=1,2,3. d) $d$-dependent THz pulse amplitude summarized from (a)–(c).

To make a better comparison, we normalized the THz pulse amplitude and the fs laser absorbance to the values obtained with the single-repeat sample with $d$=2 nm, which is the typical spintronic THz emitter of current design [28]. Both the experimental (symbols) and theoretically calculated (curves) $d$-dependent normalized quantities are depicted in **Figure 3**a–c. For the samples with n=1, these two normalized values are almost falling into an identical curve, see Figure 3a, which confirms the theoretical proposal that the THz pulse amplitude improvement originates from the fs laser absorbance enhancement. For the samples with n=2 and n=3, although all values increases as increasing *d*, the enhancement rate of the normalized THz pulse amplitude is considerably smaller than that of the normalized fs laser absorbance. And the difference becomes larger as *d* or n increases. This phenomenon can be attributed to the attenuation of the THz radiation by the periodic metal films. In fact, the stack of Pt/Fe/W films not only plays the role of THz emitter, but also attenuates the THz radiation that passes through it.

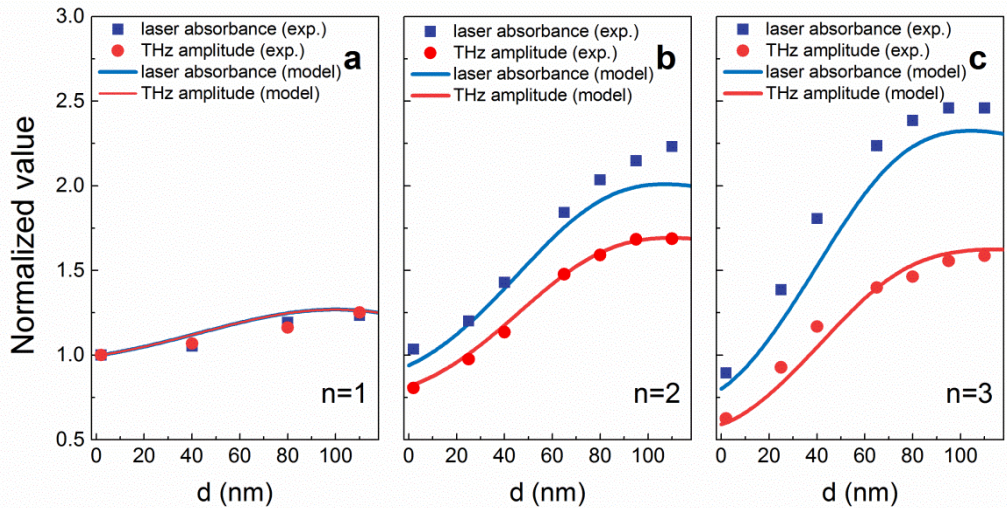

**Figure 3.** Normalized THz pulse amplitude and femtosecond laser absorbance as the functions of SiO$_2$ thickness d for different repeats: a) n=1; b) n=2; c) n=3. The reference is the sample with n=1 and *d*=2 nm. Symbols are experimental data and solid lines are the theoretically calculated results.

In the following, we establish a model to give a quantitative description. Suppose that the THz radiations from the first, second, and third periods are denoted by $E_1$, $E_2$, and $E_3$, respectively, and the corresponding laser absorbance is denoted by $A_1$, $A_2$, and

$A_3$, respectively. For the samples with n=1, according to Figure 3a, the emitted THz pulse amplitude is proportional to the laser absorbance. Such linear relation can be expressed as:

$$E_{n=1} = E_1 = \beta A_1, \tag{1}$$

where $\beta$ is a constant, denoting the THz conversion parameter. For the samples with n=2, the THz emission from the first period suffers from attenuation in the second period, while the THz radiation from the second period emits into the substrate directly. We give the expression as:

$$E_{n=2} = E_1 \cdot T_{THz} + E_2 = \beta(A_1 \cdot T_{THz} + A_2), \tag{2}$$

where $T_{THz}$ denotes THz transmittance through a period. In the same manner, the expression for the samples with n=3 can be written as:

$$E_{n=3} = E_1 \cdot T_{THz}^2 + E_2 \cdot T_{THz} + E_3 = \beta(A_1 \cdot T_{THz}^2 + A_2 \cdot T_{THz} + A_3). \tag{3}$$

Note that $A_i$ ($i$=1,2,3) can be calculated by employing transfer-matrix method (see Supporting Information Note 4), and $T_{THz}$ can be measured by THz TDS system. We have measured 12 samples with various $d$ and n, and obtained the average value of measured $T_{THz}$, $T_{THz}$=0.78. The model results are plotted as solid lines in Figure 3a–c, which show very good agreement with the experimental results.

To elucidate the physics more intuitively, we depict the laser field distribution in three samples with n=3 in **Figure 4**, whose $SiO_2$ thicknesses are 2, 40, and 110 nm, respectively. One can see that: i) the first period has the largest laser field distribution and thus can provide the strongest THz emission; ii) with the increase of $d$, the fields in different periods rise with different rate, and $A_1$ grows most rapidly. Thus the first period offers the largest contribution to THz emission enhancement. Since such emission should pass through the other two periods, the total contribution is reduced by a factor of $T_{THz}^2$, as shown in Eq. (3). The samples with n=2 share similar physics, but in contrast, the total contribution from the first period is reduced by a factor of $T_{THz}$, as shown in Eq. (2). It indicates a competition between the laser absorbance enhancement and the THz radiation attenuation while increasing the number of periods. In our experiments, the sample with n=2 and $d$=110 nm provides the strongest THz pulse emission, which presents a 1.7 times improvement compared to the single-repeat spintronic THz emitter with n=1 and $d$=2 nm.

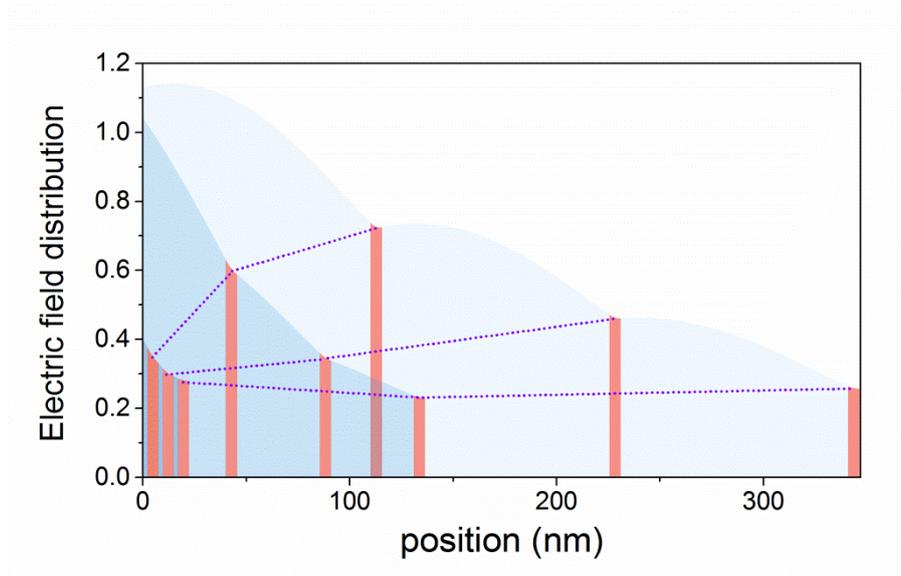

**Figure 4.** Electric field distributions in the samples [SiO$_2$($d$)/Pt(1.8 nm)/Fe(1.8 nm)/W(1.8 nm)]$_3$ with various SiO$_2$ thickness, $d$=2,40,110 nm.

In conclusion, we have demonstrated that metal-dielectric photonic crystal structure can greatly enhance the laser absorption in the NM$_1$/FM/NM$_2$ spintronic THz emitter, and thus improve the conversion efficiency. The key is that the interference between the multiple scattering waves suppresses the reflection and transmission simultaneously and reshapes the laser field distributions in the structure, which can be tailored by tuning the thickness of dielectric interlayer. Accordingly, we achieved the strongest THz pulse emission that presents a 1.7 times improvement compared to the currently designed spintronic THz emitter. A theoretical model was established to elucidate the competition between the laser absorption enhancement and the THz radiation attenuation by metal films, which showed excellent agreement with experimental results. This work introduces one of the most popular optical concepts, photonic crystal, into the burgeoning research field of spintronic THz emitter, which may inspire the researcher to exploit fruitful ideas in the future, such as plasmonics and metamaterials, etc.

## Experimental Section

*Sample Preparation*: The stack of Pt(1.8 nm)/Fe(1.8 nm)/W(1.8 nm) films and the SiO$_2$ interlayers were grown on a 0.5-mm-thick MgO substrate by dc and rf magnetron sputtering, respectively. The base pressure of the sputter chamber was $5\times 10^{-5}$ Pa.

*Experimental Details*: A standard THz time-domain spectroscopy (TDS) setup has been utilized to generate and detect the THz pulse waveforms. Linearly polarized femtosecond laser pulses (with duration of 120 fs, center wavelength of 800 nm, power of 350 mW, and repetition rate of 80 MHz) from a Ti:sapphire laser oscillator excite the emitter under normal incidence, and the generated THz signals were detected by the electro-optic sampling technique with probe pulses (20 mW) from the same laser co-propagating with the THz wave through an electro-optic crystal. A 1-mm-thick ZnTe (110) electro-optic crystal was used for detection. An in-plane magnetic field of 140 mT is applied to the emitter. The laser beam diameter was adjusted to be ~1 mm. For the optical absorption measurements, the emitters were illuminated under the same fs laser energy, and the reflected and transmitted energy were detected by power meter. All measurements were performed at room temperature in a dry air environment.

## Acknowledgements

The authors would like to thank Prof. Jingbo Qi for helpful discussions, and thank Chao Zhou, Min Meng, and Hanbin Wang for experimental assists. This work was supported by Science Challenge Project (Grant No. TZ2018003), National Natural Science Foundation of China (Grants No. 11504345, No. 11504346, No. 51571109, No. 11734006, and No. 11674379) and the National Key R&D Program of China (Grant No. 2017YFA0303202 and 2018YFA0306004), and the National Basic Research Program of China under Grant No. 2015CB921403.

## REFERENCES

[1] B. Ferguson, X. Zhang, Nat. Mater. 2002, 1, 26.

[2] R. Ulbricht, E. Hendry, J. Shan, T. Heinz, M. Bonn, Rev. Mod. Phys. 2011, 83, 543.

[3] T. Kampfrath, K. Tanaka, K. Nelson, Nat. Photon. 2013, 7, 620.

[4] L. Ho, M. Pepper, P. Taday, Nat. Photon. 2008, 2, 541.

[5] M. Hishida, K. Tanaka, Phys. Rev. Lett. 2011, 106, 158102.

[6] P. U. Jepsen, D. G. Cooke, M. Koch, Laser Photon. Rev. 2011, 5, 124.

[7] S. Koenig, D. Lopez-Diaz, J. Antes, F. Boes, R. Henneberger, A. Leuther, A. Tessmann, R. Schmogrow, D. Hillerkuss, R. Palmer, T. Zwick, C. Koos, W. Freude, O. Ambacher, J. Leuthold, I. Kallfass, Nat. Photon. 2013, 7, 977.

[8] R. Appleby, H. B. Wallace, IEEE Trans. Antennas Propag. 2007, 55, 2944.

[9] K. B. Cooper, R. J. Dengler, N. Llombart, B. Thomas, G. Chattopadhyay, P. H. Siegel, IEEE Trans. Terahertz Sci. Technol. 2011, 1, 169.

[10] M. Tonouchi, Nat. Photon. 2007, 1, 97.

[11] K. Reimann, Rep. Prog. Phys. 2007, 70, 1597.

[12] X. C. Zhang, A. Shkurinov, Y. Zhang, Nat. Photon. 2017, 11, 16.

[13] S. S. Dhillon, M. S. Vitiello, E. H. Linfield, A. G. Davies, M. C. Hoffmann, J. Booske, C. paoloni, M. Gensch, P. Weightman, G. P. Willianms, E. Castro-Camus, D. R. S. Cumming, F. Simoens, I. Escorcia-Carranza, J. Grant, S. Lucyszyn, M. Kuwata-Gonokami, K. Konishi, M. Koch, C. A. Schmuttenmaer, T. L. Cocker, R. Huber, A. G. Markelz, Z. D. Taylor, V. P. Wallace, J. A. Zeitler, J. Sibik, T. M. Korter, B. Ellison, S. Rea, P. Goldsmith, K. B. Cooper, R. Appleby, D. Pardo, P. G. Huggard, V. Krozer, H. Shams, M. Fice, C. Renaud, A. Seeds, A. Stöhr, M. Naftaly, N. Ridler, R. Clarke, J. E. Cunningham, M. B. Johnston, J. Phys. D: Appl. Phys. 2017, 50, 043001.

[14] G. Mourou, C. V. Stancampiano, A. Antonetti, A. Orszag, Appl. Phys. Lett. 1981,

39, 295.

[15] D. H. Auston, K. P. Cheung, P. R. Smith, Appl. Phys. Lett. 1984, 45, 284.

[16] Y. C. Shen, P. C. Upadhya, E. H. Linfield, H. E. Beere, A. G. Davies, Appl. Phys. Lett. 2003, 83, 3117.

[17] A. Dreyhaupt, S. Winnerl, T. Dekorsy, M. Helm, Appl. Phys. Lett. 2005, 86, 121114.

[18] C. W. Berry, N. Wang, M. R. Hashemi, M. Unlu, M. Jarrahi, Nat. Commun. 2013, 4, 1622.

[19] X. C. Zhang, Y. Jin. K. Yang, L. J. Schowalter, Phys. Rev. Lett. 1992, 69, 2303.

[20] A. Rice, Y. Jin, X. F. Ma, X. C. Zhang, D. Bliss, J. Larkin, M. Alexander, Appl. Phys. Lett. 1994, 64, 1324.

[21] R. A. Kaindl, F. Eickemeyer, M. Woerner, T. Elsaesser, Appl. Phys. Lett. 1999, 75, 1060.

[22] R. Huber, A. Brodschelm, F. Tauser, A. Leitenstorfer, Appl. Phys. Lett. 2000, 76, 3191.

[23] T. Tanabe, K. Suto, J. Nishizawa, K. Saito, T. Kimura, J. Phys. D: Appl. Phys. 2003, 36, 953.

[24] T. Löffler, M. Kreß, M. Thomson, T. Hahn, N. Hasegawa, H. G. Roskos, Semicond. Sci. Technol. 2005, 20, S134.

[25] X. Xie, J. Dai, X. C. Zhang, Phys. Rev. Lett. 2006, 96, 075005.

[26] J. Dai, X. Xie, X. C. Zhang, Phys. Rev. Lett. 2006, 97, 103903.

[27] T. Kampfrath, M. Battiato, P. Maldonado, G. Eilers, J. Nötzold, S. Mährlein, V.

Zbarsky, F. Freimuth, Y. Mokrousov, S. Blügel, M. Wolf, I. Radu, P. M. Oppeneer, M. Münzenberg, Nat. Nanotechnol. 2013, 8, 256.

[28] T. Seifert, S. Jaiswal, U. Martens, J. Hannegan, L. Braun, P. Maldonado, F. Freimuth, A. Kronenberg, J. Henrizi, I. Radu, E. Beaurepaire, Y. Mokrousov, P. M. Oppeneer, M. Jourdan, G. Jakob, D. Turchinovich, L. M. Hayden, M. Wolf, M. Münzenberg, M. Kläui, T. Kampfrath, Nat. Photon. 2016, 10, 483.

[29] D. Yang, J. Liang, C. Zhou, L. Sun, R. Zheng, S. Luo, Y. Wu, J. Qi, Adv. Opt. Mater. 2016, 4, 1944.

[30] Y. Wu, M. Elyasi, X. Qiu, M. Chen, Y. Liu, L. Ke, H. Yang, Adv. Mater. 2016, 29, 1603031.

[31] T. Seifert, S. Jaiswal, M. Sajadi, G. Jakob, S. Winnerl, M. Worf, M. Kläui, T. Kampfrath, Appl. Phys. Lett. 2017, 110, 252402.

[32] G. Torosyan, S. Keller, L. Scheuer, R. Beigang, E. Th. Papaioannou, Sci. Rep. 2018, 8, 1311.

[33] J. E. Hirsch, Phys. Rev. Lett. 1999, 83, 1834.

[34] E. Saitoh, M. Ueda, H. Miyajima, G. Tatara, Appl. Phys. Lett. 2006, 88, 182509.

[35] C. Zhou, Y. P. Liu, Z. Wang, S. J. Ma, M. W. Jia, R. Q. Wu, L. Zhou, W. Zhang, M. K. Liu, Y. Z. Wu, J. Qi, arXiv:1804.04765.

[36] C. Z. Tan, J. Non-Cryst. Solids, 1998, 223, 158.

[37] R. E. Stephens, I. H. Malitson, J. Res. Natl. Bur. Stand. 1952, 49, 249.

# Supporting Information

**Highly-efficient spintronic terahertz emitter enabled by metal-dielectric photonic crystal**

Zheng Feng[1], Rui Yu[2], Yu Zhou[3], Hai Lu[4], Wei Tan[1,*], Hu Deng[5], Quancheng Liu[5], Zhaohui Zhai[6], Liguo Zhu[6], Jianwang Cai[3,*], Bingfeng Miao[2], and Haifeng Ding[2,*]

[1] *Microsystem & Terahertz Research Center, CAEP, Chengdu 610200, China*
[2] *National Laboratory of Solid State Microstructures and Department of Physics, Nanjing University, Nanjing 210093, China*
[3] *Beijing National Laboratory for Condensed Matter Physics, Institute of Physics, Chinese Academy of Sciences, Beijing 100190, China*
[4] *Engineering Laboratory for Optoelectronic Technology and Advanced Manufacturing, Henan Normal University, Xinxiang 453007, China*
[5] *School of Information Engineering, Southwest University of Science and Technology, Mianyang 621010, China*
[6] *Institute of Fluid Physics, China Academy of Engineering Physics, Mianyang 621900, China*

*Corresponding author: *tanwei@mtrc.ac.cn, jwcai@iphy.ac.cn, hfding@nju.edu.cn*

## 1. Toy model for transmission properties of ultra-thin conducting sheet

Since the Pt(1.8 nm)/Fe(1.8 nm)/W(1.8 nm) structure is far smaller than the operating wavelength of fs laser (800 nm), we employ a toy model here that treat the film stack as a conducting sheet with a negligible thickness. Note that this method is widely used in the study of graphene [S1]. For simplicity, we consider the case that the dielectric constant of environment is uniform. Suppose that the complex conductivity of the conducting sheet is $\sigma = \sigma_r + i\sigma_i$, the transmission and reflection coefficients at normal incidence can be deduced with the assistance of Ohm's law [S1],

$$t = \frac{2}{2 + (\sigma_r + i\sigma_i)\eta_0}, \tag{S1}$$

$$r = \frac{-(\sigma_r + i\sigma_i)\eta_0}{2 + (\sigma_r + i\sigma_i)\eta_0}, \tag{S2}$$

where $\eta_0$ is the wave impedance of the free space. Then one can readily obtain the absorbance of the conducting sheet,

$$A = 1 - |r|^2 - |t|^2 = \frac{4\sigma_r \eta_0}{(2 + \sigma_r \eta_0)^2 + (\sigma_i \eta_0)^2}. \tag{S3}$$

Since $4\sigma_r \eta_0 \leq (2 + \sigma_r \eta_0)^2 / 2$, one can conclude that $A \leq 50\%$ and the maximum occurs only if $\sigma_r \eta_0 = 2$ and $\sigma_i = 0$. This is an intrinsic limit for ultra-thin conducting sheet.

It should be note here that in practical realizations, the existence of substrate makes the environment at opposite sides non-uniform, which may lead to asymmetric absorbance when light comes from different side. From one side, the absorbance may be slightly larger than 50%, while from the other side, it is less than 50%.

## 2. Transfer matrix method for one-dimensional layered structure

The transmission properties of one-dimensional layered structure can be calculated precisely by employing transfer matrix method. The electrical ($E$) and magnetic ($H$) components of the wave in the front interface of the $j$-th layer is related to those in the back interface by a transfer matrix $M_j$,

$$\begin{pmatrix} E_j \\ H_j \end{pmatrix} = M_j \begin{pmatrix} E_{j+1} \\ H_{j+1} \end{pmatrix} \tag{S4}$$

with

$$M_j = \begin{bmatrix} \cos k_j d_j & -i\eta_j \sin k_j d_j \\ -i/\eta_j \sin k_j d_j & \cos k_j d_j \end{bmatrix} \tag{S5}$$

at normal incidence. Here $k_j$ and $\eta_j$ are the wave vector and impedance in the $j$-th layer, respectively, and $d_j$ denote the thickness. The transfer matrix of the entire system can be written as

$$M = \prod_{l=1}^{N} M_l = \begin{pmatrix} m_{11} & m_{12} \\ m_{21} & m_{22} \end{pmatrix}, \quad (S6)$$

where $N$ is the number of layers. Then we can obtain the transmission and reflection coefficients,

$$t = \frac{2\eta_0}{m_{11}\eta_0 + m_{22}\eta_0 + m_{12} + m_{21}\eta_0^2} \quad (S7)$$

$$r = \frac{m_{11}\eta_0 - m_{22}\eta_0 + m_{12} - m_{21}\eta_0^2}{m_{11}\eta_0 + m_{22}\eta_0 + m_{12} + m_{21}\eta_0^2}. \quad (S8)$$

### 3. Retrieving effective permittivity of NM$_1$/FM/NM$_2$ stack

We use the standard sample SiO$_2$(2 nm)/Pt(1.8 nm)/Fe(1.8 nm)/W(1.8 nm) as a reference. The measured transmittance and reflectance of fs laser are 0.333 and 0.287, respectively. Assume that the effective permittivity is denoted by $\varepsilon_m = \varepsilon_r + i\varepsilon_i$. Then we use numerical method to fit Eqs. (S7) and (S8), and obtain that $\varepsilon_m = -34.15 + 26.08i$. In the calculations, the permittivity for SiO$_2$ interlayer is 2.13, and that for MgO substrate is 2.98.

### 4. Calculating the absorbance contribution in each period

Based on Eqs. (S7) and (S8), one can readily obtain the absorbance of the photonic crystal structure, $A = 1 - |r|^2 - |t|^2$. By further employing the transfer matrix method, we can calculate the electric field distribution in the structure, as shown in

Fig. 1(c) and Fig. 4 in the main text. Since the interlayer and the substrate are almost loss-free, the absorbance is directly proportional to the square of the electric field amplitude in the Pt/Fe/W stacks. Hence, the absorbance in each period, $A_i$ ($i$=1,2,3), can be obtained once we get the electric field amplitude. For simplicity, we compare the electric field at the center of each Pt/Fe/W stack, $E_{\text{field}}^i$ ($i$=1,2,3), to estimate the contribution of absorbance in each period. That is,

$$A_i = \frac{\left|E_{\text{field}}^i\right|^2 A}{\left|E_{\text{field}}^1\right|^2 + \left|E_{\text{field}}^2\right|^2 + \left|E_{\text{field}}^3\right|^2} \quad (i=1,2,3) . \tag{S9}$$

**Reference**

[S1] Y. Fan, F. Zhang, Q. Zhao, Z. Wei, and H. Li, "Tunable terahertz coherent perfect absorptionin a monolayer graphene," Opt. Lett. 39, 6269–6272 (2014).